\documentstyle[preprint,aps]{revtex}
\tightenlines
 
\begin{document}
\newcommand{\beq}{\begin{equation}}
\newcommand{\eeq}{\end{equation}}
\newcommand{\beqa}{\begin{eqnarray}}
\newcommand{\eeqa}{\end{eqnarray}}
\newcommand{\sr}{\sqrt}
\newcommand{\fr}{\frac}
\newcommand{\mn}{\mu \nu}
\newcommand{\G}{\Gamma}

\draft
\preprint{ INJE-TP-00-03}
\title{No Ghost State in the Brane World}
\author{Gungwon Kang\footnote{E-mail address:
kang@physics.inje.ac.kr} and Y.S. Myung\footnote{E-mail address:
ysmyung@physics.inje.ac.kr} }
\address{
Department of Physics, Inje University,
Kimhae 621-749, Korea}
\maketitle
\begin{abstract}
 
We discuss the role of the trace part of metric fluctuations 
$h_{MN}$ in the Randall-Sundrum scenario of gravity. 
Without the matter, this field ($h=\eta^{MN}h_{MN}$) is a
gauge-dependent term, and thus it can be gauged away. But, 
including the uniform source $\tilde{T}_{MN}$, this field 
satisfies the linearized equation $\Box_4 h =16\pi G_5 
T^{\mu}_{\mu}$. This may correspond to the scalar $\xi^5$
in the bending of the brane due to the localized source. 
Considering the case of longitudinal perturbations 
($h_{5\mu} =h_{55}=0$), one finds the source relation 
$\tilde{T}^{\mu}_{\mu}=2\tilde{T}_{55}$, which leads to 
the ghost states in the massive modes. In addition, if one
requires $T_{44}=2(T_{22}+T_{33})$, it is found that in the limit
of $m^2_h \to 0$  we have the massless spin-2 propagation without
the ghost state. This exactly corresponds to the same situation 
as in the intermediate scales of Gregory-Rubakov-Sibiryakov (GRS) 
model.   
 
\end{abstract}
\bigskip
 
\newpage
 
\section{Introduction}
 
Recently, there have been lots of interest in the phenomenon
of localization of gravity proposed by Randall and Sundrum
(RS)~\cite{RS} (for previous relevant work see references 
therein). RS assume a single positive tension 3-brane and 
a negative bulk cosmological constant in the five dimensional
spacetime. By considering metric fluctuations from a background
which is isomorphic to sections of $AdS_5$, they have shown that
it reproduces the effect of four dimensional gravity localized
on the brane without the need to compactify the extra dimension
due to the ``warping" in the fifth dimensional space. 
In more detail, the solution to linearized equations 
in the five dimensions results in a zero mode, which can be 
identified with the four dimensional massless graviton, and 
the massive continuum Kaluza-Klein (KK) modes.   
Surprisingly, the wavefunctions of the massive continuum 
KK modes are suppressed at the brane for small energies, and 
thus ordinary gravity localized on the brane is 
reproduced at large distances.      

On the other hand, Gregory, Rubakov and Sibiryakov (GRS)~\cite{GRS}
have recently considered a brane model which is not asymptotically 
$AdS_5$, but Minkowski flat. In the GRS model, however, the ordinary
4D Newton potential is reproduced at intermediate scales only not
because of the massless zero mode, but for the resonance of zero
mass in the continuum KK spectrum~\cite{GRS,CEH,DGP,Witten}.   
In Ref.~\cite{DGP}, however, it is pointed out that the 
$m_h \rightarrow 0$ limit of a massive graviton propagator does not 
reproduce the massless graviton propagator due to the missmatch 
of the number of polarizations. In this sense the GRS model of 
``quasi-localization" of gravity would differ from the RS model. 
Contrary to it, Cs\'aki, Erlich and Hollowood~\cite{CEH2} recently
have argued that in the presence of localized source at $z=0$ 
the bending of the brane exactly compensates for the effects of 
the extra polarization in the massive graviton propagator. 
Thus the graviton propagator 
at intermediate scales is equivalent to the massless propagator of the 
Einstein theory just as in the RS scenario. However, at ultra large
scales this effective theory includes scalar anti-gravity~\cite{GRS2}.
This problem may be cured by the RG analysis~\cite{CEHT}. Also the
authors in Ref.~\cite{DGP2} point out that the mechanism to cancel 
the unwanted extra polarization leads to the presence of ghost. 
In order to have a well-defined theory, the ghost should 
disappear.   

In this paper, we investigate the non-traceless metric fluctuations
in the presence of uniform source along $z$-axis in the RS model. 
We introduce the trace field ($h$) here instead of $\xi^5$ in 
Ref.~\cite{GT}. It shows that massive graviton modes contain ghost 
states which can be removed by assuming a further condition on the 
matter source. Our work corresponds to an alternative realization 
of the results in Ref.~\cite{CEH2}.

\section{Linearized perturbations}
 
The Randall-Sundrum model with a single domain wall 
(or brane)~\cite{RS} perpendicular to the infinite fifth 
direction can be described by the following 
action:
\beq
I = \int d^4x \int^{\infty}_{-\infty} dz \Big [
    \fr{1}{16\pi G_5} \sr{-\hat{g}} (\hat{R} -2\Lambda )
    -\sr{-\hat{g}_B} \sigma (z) +{\cal L}_M  \Big ] .
\eeq
Here $G_5$ is the five dimensional Newton's constant,
$\Lambda$ the bulk cosmological constant of five dimensinal
spacetime, $\hat{g}_B$ the determinant of the metric describing
the brane, and $\sigma (z)=\sigma \delta (z)$, $\sigma$ the tension 
of the brane. $I_M = \int d^4xdz {\cal L}_M$ denotes the matter 
action, and it contributes only in the linearized level. 
In this paper, we use the signature $(-, +, +, +, +)$. 
 
If we introduce a conformal factor as follows
\beq
ds^2 = \hat{g}_{MN} dx^Mdx^N = H^{-2} g_{MN} dx^Mdx^N ,
\eeq
the field equation becomes~\cite{IV}
\beqa
\mbox{} & & G_{MN} +3\fr{\nabla_M\nabla_N H}{H} -3g_{MN}
\Big [ \fr{\nabla_P\nabla^P H}{H} 
-2\fr{\nabla_P H \nabla^P H}{H^2} \Big ] \nonumber   \\
& & = 8\pi G_5 \Big [ -\fr{\Lambda}{8\pi G_5H^2} g_{MN}
-\fr{\sr{-g_B}}{\sr{-g}} \fr{|H|}{H^2} \sigma (z) g_{\mu\nu}
\delta^{\mu}_M \delta^{\nu}_N -\fr{2}{\sr{-\hat{g}}}
\fr{\delta I_M}{\delta \hat{g}^{MN}} \Big ] 
\label{FE0}
\eeqa
with the Einstein tensor $G_{MN}$ constructed from the metric
$g_{MN}$. Now it is straightforward to see that, 
in the absence of matter source except for the domain wall 
itself (i.e., $\delta I_M/ \delta \hat{g}^{MN} =0$), the most 
general solution having a four dimensional Poincar\'e 
symmetry is 
\beq
ds^2 = H^{-2}(z) (\eta_{\mu\nu}dx^{\mu}dx^{\nu} +dz^2),
\eeq
where $H= k|z|+1$, $\Lambda =-6k^2 (< 0)$, and 
$\sigma =3k/4\pi G_5$. 

Let us consider metric fluctuations around this background 
spacetime as follows~\cite{IV,MKL}:  
\beq
g_{MN} = \eta_{MN} +h_{MN}.
\eeq
Defining $\bar{h}_{MN} =h_{MN}-\fr{1}{2}\eta_{MN} h$ where 
$h=\eta^{MN}h_{MN}$, the linearized perturbation equation of
Eq.~(\ref{FE0}) is  
\beqa
\mbox{} && -\fr{1}{2} \Box \bar{h}_{MN} +\partial_{(M}
\partial^P\bar{h}_{N)P} -\fr{1}{2} \eta_{MN} \partial^P
\partial^Q \bar{h}_{PQ} -\fr{3\partial^PH}{2H}(\partial_M
h_{NP} +\partial_N h_{MP} -\partial_P h_{MN}) \nonumber \\
&& -3\eta_{MN} \Big [ \Big (-\fr{\partial^P\partial^QH}{H}  
+2\fr{\partial^PH\partial^QH}{H^2}\Big )h_{PQ} 
-\fr{\partial^QH}{H} \partial^P \bar{h}_{PQ} \Big ] 
-3 \Big (\fr{\Box H}{H} -2\fr{\partial_PH
\partial^P H}{H^2}\Big ) h_{MN}  \nonumber  \\
&& +8\pi G_5 H^{-2}\Bigg \{ \fr{\Lambda}{8\pi G_5} h_{MN} 
+|H| \sigma \delta (z) 
\Big [ \fr{1}{2}(\eta^{\alpha\beta}h_{\alpha\beta}  
-\eta^{PQ}h_{PQ}) \eta_{\mu\nu}\delta^{\mu}_M\delta^{\nu}_N 
+\delta^{\mu}_M\delta^{\nu}_N h_{\mu\nu} \Big ] 
\Bigg \}  \nonumber  \\
&=& 8\pi G_5 \tilde{T}_{MN}, 
\label{LFE0}
\eeqa
where the linearized five dimensional matter source 
$\tilde{T}_{MN}=-\delta (2\delta I_M /\sr{-\hat{g}} \delta 
\hat{g}^{MN})$ is included, 
and $\Box =\eta^{MN} \partial_M \partial_N$. 

Taking the harmonic gauge condition,
\beq
\partial^M \bar{h}_{MN} =0 \qquad\qquad {\rm or} \qquad 
\qquad   \partial^M h_{MN} =\fr{1}{2} \partial_N h ,
\label{HG}
\eeq
the linearized field equation becomes
\beqa
\mbox{} && \Box \bar{h}_{MN} +3\fr{\partial_5H}{H} (\partial_M
h_{5N} +\partial_N h_{5M} -\partial_5 h_{MN}) +\eta_{MN} 
\fr{12k^2}{H^2} h_{55}    \nonumber  \\
&& +\fr{12k}{H} \delta (z) \Big [ (h_{MN} -\eta_{MN} h_{55}) 
-(h_{\mu\nu}-\fr{1}{2}\eta_{\mu\nu}h_{55}) \delta^{\mu}_M
\delta^{\nu}_N \Big ] = -16\pi G_5 \tilde{T}_{MN} . 
\label{LFE1}
\eeqa
In the case of $h=0$ and $\tilde{T}_{MN}=0$, this leads to 
the equation (5) in Ref.~\cite{MKL}. In components, it becomes
\beqa
\mbox{} && (\Box +3\fr{\partial_5H}{H} \partial_5 )\bar{h}_{55}
+\fr{12k^2}{H^2} \bar{h}_{55} -(\fr{\partial_5H}{H}\partial_5 
+\fr{4k^2}{H^2}) \bar{h} = -16\pi G_5 \tilde{T}_{55},  \\ 
\label{h55} 
&& (\Box +\fr{12k}{H} \delta (z)) \bar{h}_{5\mu} +3\fr{\partial_5H}
{H}\partial_{\mu} \bar{h}_{55} -\fr{\partial_5H}{H} \partial_{\mu}
\bar{h} = -16\pi G_5 \tilde{T}_{5\mu},   \label{h5m}   \\
&& (\Box -3\fr{\partial_5H}{H}\partial_5 )\bar{h}_{\mu\nu} +3\fr{
\partial_5H}{H}(\partial_{\mu}\bar{h}_{5\nu} +\partial_{\nu}\bar{h}
_{5\mu}) +\eta_{\mu\nu}(\fr{12k^2}{H^2} -\fr{6k}{H}\delta (z))\bar{h}
_{55}    \nonumber   \\
&& +\eta_{\mu\nu} (\fr{\partial_5H}{H} \partial_5 -\fr{4k^2}{H^2} 
+\fr{2k}{H} \delta (z)) \bar{h} = -16\pi G_5 \tilde{T}_{\mu\nu} , 
\label{hmn}
\eeqa
where $\bar{h}=\eta^{MN}\bar{h}_{MN}=-\fr{3}{2}h$. 
 
For longitudinal metric fluctuations (i.e., $h_{55} = h_{5\mu}=0$),
one has 
\beq
h=\eta^{\mu\nu}h_{\mu\nu}+h_{55}=h^{\mu}_{\mu}=-\bar{h}^{\mu}_{\mu},
\qquad \bar{h}_{55} =-\fr{1}{2}h^{\mu}_{\mu}, 
\qquad \bar{h}_{5\mu}=0, \qquad \bar{h} =3\bar{h}_{55}. 
\eeq
The harmonic gauge condition in Eq.~(\ref{HG}) gives 
\beq
\partial^{\mu}\bar{h}_{\mu\nu} =0, \qquad\qquad 
\partial_5 \bar{h}_{55} =0.
\label{HG2}
\eeq
Consequently, we have $\partial_5\bar{h}=\partial_5h= 
\partial_5 h^{\mu}_{\mu}=\partial_5 \bar{h}^{\mu}_{\mu}=0$.
Using these results, we find from Eq.~(\ref{h5m}) that 
$\tilde{T}_{5\mu}=0$. And other two linearized equations become
\beqa
\mbox{} && \Box \bar{h}_{55} = -16\pi G_5 \tilde{T}_{55},  
\label{h552} \\
&& (\Box -3\fr{\partial_5H}{H}\partial_5)\bar{h}_{\mu\nu}
=-16\pi G_5 \tilde{T}_{\mu\nu}.
\label{hmn2} 
\eeqa
Acting $\partial^{\mu}$ on Eq.~(\ref{hmn2}) and using the gauge 
condition Eq.~(\ref{HG2}) lead to the source conservation law
\beq
\partial^{\mu}\tilde{T}_{\mu\nu}=0.   
\label{CL}
\eeq
This is a relic of the 4D general covariance on the brane. 
By taking the trace of Eq.~(\ref{hmn2}), we also find
\beq
\Box_4 h^{\mu}_{\mu} = 16\pi G_5 \tilde{T}^{\mu}_{\mu} \qquad
\qquad   {\rm with} \qquad \Box_4 =\eta^{\mu\nu}\partial_{\mu}
\partial_{\nu} .  
\label{Trace}
\eeq
This means that the trace can propagate on the brane if one 
includes the matter source. Note, however, this corresponds to
a massless scalar propagation. Considering $h^{\mu}_{\mu}
=-2\bar{h}_{55}$, the consistency between Eq.~(\ref{h552}) 
and Eq.~(\ref{Trace}) requires the following relation 
\beq
\tilde{T}_{55} = \fr{1}{2}\tilde{T}^{\mu}_{\mu}.
\label{tr55}
\eeq
This is exactly the stabilization condition implemented in 
Refs.~\cite{KKOP,KR}. From Eqs.~(\ref{HG2}) and (\ref{Trace}) 
one obtains additional constraints as
\beq
\partial_5 \tilde{T}^{\mu}_{\mu} = \partial_5 \tilde{T}_{55} =0.
\label{constT}
\eeq

For our purpose, let us choose here the uniform source along 
$z$-axis  
\beq
\tilde{T}_{MN} = 
\left(\begin{array}{cc}
\fr{T_{\mu\nu}(x)}{L}&0\\
0&\fr{T_{55}(x)}{L}
\end{array}\right), 
\label{source}
\eeq
which satisfies Eq.~(\ref{constT}). 
Here the size $L$ of the extra space is still finite
as is shown by
\beq
L=2\int^{\infty}_{0} \sqrt{g_{55}}dz = 2\int^{\infty}_{0}
\fr{dz}{kz+1} = \fr{2}{k}\ln [kz+1]|^{\infty}_{0} 
\propto \fr{1}{k} .
\eeq
We note that $\ln [k\cdot \infty +1] $ is still finite but 
it is very small compared with $1/k$. This is because 
$k$ is allowed for up to very small quantity as the Plank
scale ($10^{-34}$ cm). 
Then we find 
\beq
8\pi G_5 \tilde{T}_{MN} = 8\pi G  
\left(\begin{array}{cc}
T_{\mu\nu}(x)&0\\
0&T_{55}(x)
\end{array}\right), 
\eeq
where $G=G_5/L$ is the four dimensional Newton's constant.

Using Eq.~(\ref{Trace}), Eq.~(\ref{hmn2}) takes the form
\beq
(\Box_4 -m^2_h)h_{\mu\nu} =-16\pi G_5 (\tilde{T}_{\mu\nu} 
-\fr{1}{2} \eta_{\mu\nu} \tilde{T}),
\label{hmn3}
\eeq
where $\tilde{T}=\eta^{\rho\sigma}\tilde{T}_{\rho\sigma}
=\tilde{T}^{\rho}_{\rho}$. Here the mass squared $m^2_h$ 
is defined by the Schr\"{o}dinger-like equation 
\beq
\big [ -\fr{1}{2}\partial^2_5 +\fr{15k^2}{8H^2} -\fr{3k}{2H}
\delta (z) \big ] \psi (z) = \fr{1}{2}m^2_h \psi (z)
\eeq
with $h_{\mu\nu}(x,z) = H^{3/2}\psi (z) \hat{h}_{\mu\nu}(x)$. 

Now we examine the graviton propagator on the brane at
$z=0$ by considering only $h_{\mu\nu}(x,0) \sim 
\hat{h}_{\mu\nu} (x)$, which satisfies
\beq
(\Box_4 -m^2_h)\hat{h}_{\mu\nu}
=-16\pi G (T_{\mu\nu} -\fr{1}{2} \eta_{\mu\nu} T) .
\eeq
It requires the bilinear forms of the source with the inverse
propagator to isolate the physical modes. As the present analysis
is on the classical level, we express $\hat{h}_{\mu\nu}$ 
in terms of source. Taking Fourier transformation to momentum
space results in
\beq
\hat{h}_{\mu\nu} (p) = \fr{16\pi G}{p^2+m^2_h} \Bigg [ 
T_{\mu\nu}(p) -\fr{1}{2}\eta_{\mu\nu} T(p) \Bigg ] .
\eeq
Then the one graviton exchange amplitude for the source
$T_{\mu\nu}$ is given by~\cite{MKL,KR}
\beq
A^{\rm class} = \fr{1}{4} \hat{h}_{\mu\nu}(p)
T^{\mu\nu}(p) = \fr{4\pi G} {p^2+m^2_h} 
(T^{\mu\nu}T_{\mu\nu} -\fr{1}{2} T^2). 
\label{sgea}
\eeq

In order to study the massive states, it is best to use 
the rest frame~\cite{SS} in which
\beq
p_1 \neq 0, \qquad\qquad  p_2 =p_3 =p_4 =0. 
\label{massivefr}
\eeq
Considering Eqs.~(\ref{CL}) and (\ref{massivefr}) leads to
the following source relations
\beq
T_{11} =T_{12} =T_{13} =T_{14} =0.
\eeq
Thus, one obtains
\beq
T^{\mu\nu}T_{\mu\nu} -\fr{1}{2}T^2 =|T_{+2}|^2 +|T_{-2}|^2 
+|T_{+1}|^2 +|T_{-1}|^2 +T_{44}\big [ \fr{1}{2}T_{44}  
-(T_{22}+T_{33}) \big ],
\label{sgeam} 
\eeq
where the first two terms correspond to the exchange of graviton
with $T_{\pm 2}=\fr{1}{2}(T_{22}-T_{33}) \pm iT_{23}$, and
the third and fourth terms are the exchange of the graviphoton 
with $T_{\pm 1}=T_{24} \pm iT_{34}$. 
We note here that the last term in the above equation is
{\it not} positive definite. This means that there exist
ghost states (negative norm states). However, if one requires
\beq
T_{44} =2(T_{22}+T_{33}),
\label{gf}
\eeq
one immediately finds that
\beq
T^{\mu\nu}T_{\mu\nu} -\fr{1}{2}T^2 =|T_{+2}|^2 +|T_{-2}|^2
+|T_{+1}|^2 +|T_{-1}|^2
\eeq
with all positive norm states. 

In the limit of $m^2_h \rightarrow 0$, the graviphoton propagation
can be decoupled from the brane~\cite{vVZ}. Hence we can neglect 
$|T_{\pm 1}|^2$-terms. Finally the amplitude takes the form
\beq
A^{\rm class}_{m^2_h \rightarrow 0} = \lim_{m^2_h \rightarrow 0} 
\fr{4\pi G}{p^2_1 +m^2_h} 
\big [ |T_{+2}|^2 +|T_{-2}|^2 \big ] , 
\label{sgea0}
\eeq
which corresponds to the massless spin-2 amplitude.
This is our key result. Although this is based on the second 
RS model, the results obtained above are directly applicable
to the situation in the intermediate scales of the GRS 
model~\cite{GRS,MK}.

\section{Discussion}

We resolve the problem raised in the mechanism to cancel 
the unwanted extra polarization in the quasi-localization of 
gravity. This is done with introducing both the trace ($h$) and
the uniform source ($\tilde{T}_{MN}$) at the linearized level. 
In the conventional RS approach, the trace ($h$) is just 
a gauge-dependent scalar and hence it can be gauged away. 
However, including the uniform matter source, this plays 
the role of $\xi^5$ in the brane bending model~\cite{GT}.
This is because $h$ ($\xi^5$) satisfy the nearly same massless
equations of $\Box_4 h = 16\pi G_5 \tilde{T}^{\mu}_{\mu}$ 
($\Box_4 \xi^5 = \fr{8\pi G_5}{6} S^{\mu}_{\mu}$ in 
Ref.~\cite{GT}). And the comparison of the equation 
$\bar{h}_{\mu\nu} =h_{\mu\nu}-\fr{1}{2}\eta_{\mu\nu} h$ with 
$\bar{h}_{\mu\nu} =h^{(m)}_{\mu\nu} +2k \eta_{\mu\nu}\xi^5$
in Ref.~\cite{GT} confirms the close relationship between $h$ 
and $\xi^5$. 

If $T^{\mu}_{\mu}=0$, one finds from Eq.~(\ref{sgea}) that the 
massive spin-2 states have 5 polarizations with all positive 
norm states~\cite{MKL}. In the case of $h \neq 0$,  
$T^{\mu}_{\mu} \neq 0$, requiring the additional 
condition $T_{44} =2(T_{22}+T_{33})$ in Eq.~(\ref{gf}), 
we find the massless spin-2 state with 2 polarizations in the
limit of $m^2_h \to 0$. In this case the ghost states disappear. 

Now we wish to comment a recent paper 
by Kogan and Ross~\cite{KR}. They require only 
$\tilde{T}_{55}=\fr{1}{2}
\tilde{T}^{\mu}_{\mu}$ in Eq.~(\ref{tr55}). This condition can 
be interpreted as follows: $\tilde{T}^{\mu}_{\mu}-2\tilde{T}_{55}$
is the source for the scalar radion. In the case of a mechanism 
that stabilizes the extra dimension, the source for the constant 
mode of this scalar is identically zero. 
In our case this comes from the consistency
between Eqs.~(\ref{h552}) and (\ref{Trace}) with 
$h^{\mu}_{\mu}=-2\bar{h}_{55}$.
In Ref.~\cite{KR}, Kogan and Ross neither choose the massive frame 
of Eq.~(\ref{massivefr}) nor use the source conservation law 
in Eq.~(\ref{CL}). These two are essential steps for obtaining 
the massive spin-2 amplitude of Eq.~(\ref{sgea0}). In the massless
frame of $p_1=p_4$, $p_2=p_3=0$ with $p^{\mu}T_{\mu\nu}=0$ 
\cite{SS,Wein}, $T^{\mu\nu}T_{\mu\nu} -\fr{1}{2}T^2 $ 
exactly reduces to $|T_{+2}|^2 +|T_{-2}|^2$ without the ghost 
states. But in the massive frame it leads to 
Eq.~(\ref{sgeam}). Hence one finds the ghost states. To eliminate
these, we require a further condition as Eq.~(\ref{gf}).   

We are still under the harmonic gauge in Eq.~(\ref{HG}). 
We note that the expression for $A^{\rm class}$ is simplified
by refering it to an appropriate Lorentz frame. Usually one 
chooses the rest frame $p_{\mu}=(p,0,0,0)$ to see the massive 
states~\cite{Velt}. On the other hand, we use the light-cone 
frame of $p_{\mu}=(p,0,0,p)$ for the massless states. 
The ghost-free condition requires
a further relation among diagonal elements of $T_{MN}$.
We are free from the ghost provided that $T^{\mu}_{\mu}
=3(T_{22}+T_{33})=2T_{55}$.

Finally, we comment on our source $T_{MN}$ 
in Eq.~(\ref{source}). In the brane-bending approach~\cite{GT},
the authors choose a localized source on the brane 
as $T_{\mu\nu}(x,z) =T_{\mu\nu}(x)\delta (z)$. 
Here we choose $\tilde{T}_{\mu\nu}(x,z)
= T_{\mu\nu}(x)/L$. But, up to the integration over $z$, 
these two expressions lead to the same one. In the case of 
``brane-bending," the source is located on the brane at 
$z=0$ whereas our source is uniformly distributed along
the extra dimension $z$.

\section*{Acknowledgments}

The authors thank Hyungwon Lee for helpful discussions.  
This work was supported by the Brain Korea 21 Programme, Ministry of
Education, Project No. D-0025.

\end{document}